\documentclass{aa}  
\usepackage{tabularx} 
\usepackage{adjustbox}
\usepackage[dvipsnames]{xcolor}
\usepackage{ulem}

\usepackage{graphicx}
\usepackage{txfonts}
\usepackage{hyperref}
\begin{document}

   \title{Galaxy sizes and compactness at Cosmic Dawn}

    \author{ P. Cataldi\inst{1},  S. Pedrosa\inst{1},  L.J. Pellizza\inst{1}\fnmsep\thanks{Corresponding author.}, D. Ceverino\inst{2,3}, L.A. Bignone\inst{1}
}

\institute{
Instituto de Astronom{\'\i}a y F{\'\i}sica del Espacio, CONICET--UBA, Argentina
\and Departamento de Fisica Teorica, Universidad Autonoma de Madrid, Madrid, Spain
\and CIAFF, Facultad de Ciencias, Universidad Autonoma de Madrid, 28049 Madrid, Spain}

   \date{Received --; accepted --}

 
  \abstract
  {The \textit{James Webb Space Telescope} has found an unexpected population of high-mass galaxies ($\log (M^\star / \mathrm{M}_\odot) \gtrsim 10$) with extremely small effective radii ($\sim 100\,\mathrm{pc}$) at $z \gtrsim 6$. Also, the existence of an unusual size--mass relation has been claimed. These observations are only partially reproduced by current models, and the physics responsible for the observed relations is still under debate.}
  {We aim at understanding the physical mechanisms governing the size evolution of galaxies, and its dependence on their properties in the Early Universe. We expect to unveil the formation channels of the observed compact galaxies.}
  {We analyse 7605 snapshots for 169 galaxies of the state-of-the-art cosmological simulation suite \textsc{FirstLight}, focusing on the high-redshift stellar size--mass relation and its evolution with a resolution of tens of parsecs.}
  {We find that galaxies undergo an expansion--compaction--re-expansion process. The sizes attained by galaxies during compaction are comparable with those observed. This process operates in a specific mass range; compaction starts at $\log (M^\star_\mathrm{on} / \mathrm{M}_\odot) \sim 8.5$ and ends at $\log (M^\star_\mathrm{off} / \mathrm{M}_\odot) \sim 9.5$. In between these masses, the size--mass relation becomes inverted, with a negative slope. The physical mechanism driving this process in our simulations involves a self-reinforced inflow of gas from the outer regions, that triggers a strong, localised starburst at the centre (within $1\,\mathrm{kpc}$). This contraction continues until conditions favour star formation in a broader area, and the normal inside-out growth pattern resumes.}
  {We present evidence for the existence of a universal wet compaction operating at Cosmic Dawn. This mechanism is driven by spherical accretion triggered by the change of the state of the central matter of galaxies, from dark matter- to baryon-dominated. We also propose an analytical expression for the infall process, suitable for use in semi-analytic models. Contrary to low-redshift galaxies, in high-redshift systems compaction ends without gas depletion and star-formation quenching.}

   \maketitle
%

\section{Introduction}

Sizes and internal kinematics constitute fundamental proxies for understanding the processes involved in the formation of galaxies, and those shaping their structures. These proxies are the result of many internal and external multi-scale physical mechanisms, such as galaxy mergers, instabilities, gas accretion and outflows, feedback processes, and star formation. However, the precise connection between the size and kinematics of the stellar component and these mechanisms is still poorly known \citep[e.g.][and references therein]{Hopkins2023}.

The Sloan Digital Sky Survey has provided a large statistical sample of galaxies in the Local Universe, measuring reliable data on their properties \citep[e.g.][]{Kauffmann2003}. These data, together with those from the \textit{Hubble Space Telescope} (HST), allowed to investigate fundamental relations that help to unveil the aforementioned connection, such as that between stellar size and mass \citep{Shen2003,Simard2011,Cappellari2013,vanderWel2014,Mowla2019b,Mowla2019a}. This relation presents a positive slope at low redshifts ($z \lesssim 2$), indicating that galaxies grow inside-out.

The picture is not so clear at higher redshifts.
The HST has revealed a population of massive galaxies ($M^\star \sim 10^{10}\,\mathrm{M}_{\odot}$) at $z \sim 7$, that are more compact than their counterparts at lower redshifts \citep[e.g,][]{Bruce2012,vanderWel2014,Allen2017,Bouwens2017,Yang2021}. The \textit{James Webb Space Telescope} (JWST) has unveiled an unexpected population of red galaxies at $z \sim 7 - 9$, with stellar masses $M^\star \gtrsim 10^{10} \mathrm{M}_\odot$, extremely small effective radii ($\left< r_\mathrm{e} \right> \approx 150 \, \mathrm{pc}$), high star formation rate surface densities ($\Sigma_\mathrm{SFR} > 10\, \mathrm{M}_\odot \, \mathrm{yr}^{-1} \, \mathrm{kpc}^{-2}$) and positive evidence of inside-out growth  \citep{Baggen2023,Baker2024,Morishita2024,Ormerod2024,Ward2024,Miller2025,Yang2025}. Also \citet{Tacchella2024} have found compact galaxies with sizes of the order of $100\,\mathrm{pc}$ in the JADES survey. Interestingly, \citet{Baggen2023} report an inverted (i.e. negative slope) stellar size--mass relation, whereas \citet{Morishita2024} argue that, while the stellar size--UV luminosity relation is inverted, the size--mass relation is normal. The question of how massive galaxies become compact and the inverted size--mass relation (if confirmed) arises, should therefore be addressed.

Theoretical models suggest that the normal size--mass relation arises because the gas infalling to the central regions conserves its angular momentum, forming a disk that grows in an inside-out way \citep{Kravtsov1997,Mo1998}. Star formation proceeds therefore at larger radii as the mass of the system increases. High-resolution cosmological hydrodynamical simulations confirm this picture \citep{Genel2014,Pedrosa2015,Pillepich2018,Popping2022}, and are indeed are capable of reproducing the relation at low redshifts \citep{Furlong2017}.

Challenges appear in galaxies at high redshifts. Numerical studies of galaxies at the Cosmic Dawn, with different baryonic implementation and numerical resolution \citep[\textsc{BlueTides, flares ix, Thesan;}][]{Ni2020,Lovell2021,Kannan2022}, obtain an intrinsic size--mass relation with a maximum at a stellar mass between $10^8$ to $10^9\,\mathrm{M}_\odot$. At lower masses galaxies are influenced by the strength of feedback-driven outflows, resulting in growing sizes as mass increases. In contrast, massive galaxies are affected by disk instability that triggers compaction (i.e., a decrease in the size of the stellar distribution\footnote{As measured by, e.g., its half-mass radius.}), inverting the relation. Also \citep{Roper2023} find a transition between extended and irregular objects to compact ones via efficient centralised cooling that produces a high star formation rate (SFR) in their cores. However, they find a normal size--mass relation and argue that the observed inversion of the size--luminosity trend is due to the effect of dust.

Several works explore the mechanisms behind size changes, either theoretically \citep{Dekel2014} or using cosmological simulations \citep{Zolotov2015,Lapiner2024}. They present different versions of the ``wet compaction'' scenario, in which the dissipative collapse of high-redshift, gas-rich, turbulent discs produces compact star-forming systems (``blue nuggets'') at $z \lesssim 4-5$. These discs are fed by cold streams or mergers, and undergo instabilities that drive gas inward. Compaction triggers central gas depletion and quenching, making galaxies to evolve into compact spheroids (``red nuggets'') by $z \sim 2$. Quenching proceeds inside-out, thus galaxies resume expansion as star formation moves outwards into a ring \citep{Dekel2020}. At low redshift, blue nuggets are rare due to reduced gas content. According to \citet{Lapiner2024}, wet compaction occurs near a critical stellar mass of $\sim 10^{10}\, M_\odot$. Driven by angular momentum loss, this transition reshapes their structure, composition, and kinematics --- from diffuse to compact, dark matter to baryon central dominance, and supernova- to AGN-driven feedback. Also, \citet{Shen2024} find that low-mass galaxies ($M^\star \sim 10^{7-9}\,\mathrm{M}_\odot$) at $z > 3$ do not undergo a compaction--re-expansion process, but rapid size fluctuations produced by the competition between feedback-driven gas outflows and cold inflows. These fluctuations preserve the normal size--mass relation, enlarging its dispersion instead of changing its slope.

Given the dissimilar results about the high-redshift size--mass/luminosity relation found by both observations and various simulations, a thorough investigation is necessary to elucidate the involved processes. However, numerical resolution of a few hundred parsecs from large-scale cosmological simulations such as \textsc{IllustrisTNG} \citep{Pillepich2018}, \textsc{Thesan} \citep{Shen2024}, \textsc{flares} \citep{Roper2022}, \textsc{BlueTides} \citep{Marshall2022} and \textsc{simba} \citep{Wu2020}, makes hard to resolve internal structures at high redshift.

High-resolution simulations such as \textsc{Thesan-zoom} \citep{McClymont2025} or \textsc{FIREbox} \citep{Feldmann2023} are the present workhorses used to overcome this limitation.
In this paper, we use the \textsc{FirstLight} (FL) simulation suite \citep{Ceverino2017,Ceverino2018,Ceverino2019,Ceverino2021,Ceverino2024}, which achieves a high spatial resolution of $\sim 10\, \mathrm{pc}$ and provides a large sample size for a robust analysis. We used snapshots of 169 simulated galaxies, each at $45$ redshifts in the range $5.25 - 9$ (a total of $7605$ snapshots). We aim at studying the growth of the stellar distribution of galaxies at Cosmic Dawn, and what are the physical mechanisms responsible for the high-mass galaxy compaction and the inversion of the mass--size relation.

The structure of this paper is as follows. Sect.~\ref{simulations} provides a brief synopsis of the FL simulations used throughout this work. Sect.~\ref{results} describes our results on the evolution of the sizes of galaxies, whereas Sect.~\ref{conclusions} compares them to those obtained by other works, discusses their relevance to the understanding of galaxy evolution in the Early Universe, and presents our conclusions.


\section{The simulation}
\label{simulations}


This paper uses a subsample from the cosmological zoom-in simulation suite described in detail in \citet{Ceverino2017}. The FL project has been capable of predicting the mechanisms driving galaxy morphology \citep{Nakazato2024,Ceverino2021}. They have also produced a galaxy population that matches a wide range of observations at high redshifts such as star-formation histories \citep{Ceverino2018,Ceverino2024}, rest-frame UV/optical absolute magnitudes and colours, mass-metallicity \citep{Langan2020} and mass-dust relations \citep{Mushtaq2023}, and optical emission lines \citep{Ceverino2019,Ceverino2021,Nakazato2023}. 

The simulations are performed with the \textsc{ART} code \citep{Kravtsov1997, Kravtsov2003,Ceverino2009}, which accurately follows the evolution of a gravitating \textit{N}-body system and the gas dynamics using an Eulerian adaptive mesh refinement (AMR) approach. Besides gravity and hydrodynamics, the code incorporates many astrophysical processes relevant to galaxy formation as subgrid physics that include gas cooling due to atomic hydrogen and helium, metal and molecular hydrogen cooling, photoionization heating by a constant cosmological UV background with partial self-shielding \citep{Haardt1996}, star formation and feedback (thermal, kinetic, and radiative). 

Stellar particles are created in gas cells at numerical densities above a threshold of $1\, \mathrm{cm}^{-3}$ and temperatures below $10^4\,\mathrm{K}$. The code implements a stochastic star formation model that yields the empirical Kennicutt--Schmidt law \citep{Schmidt1959, Kennicutt1998,Ceverino2009}. The simulations track the metals released from SNe Ia and SNe II, using yields from \citet{Woosley1995}.

In addition to thermal energy feedback the simulations use radiative feedback, modelled as a non-thermal pressure added to the total gas pressure in regions where ionizing photons from massive stars are produced and trapped. It uses moderate trapping of infrared photons \citep[details in][]{Ceverino2014}. The code also includes the injection of momentum from the (unresolved) expansion of gaseous shells from SNe and stellar winds \citep{Ostriker2011}. More details can be found in \citet{Ceverino2017}, \citet{Ceverino2009}, \citet{Ceverino2010}, and \citet{Ceverino2014}. The effect of the different feedback models is discussed in \citet{Ceverino2014} and \citet{Ceverino2023}. AGN feedback is not included in these simulations. It may have an effect on quenching SF at very high stellar masses, $M^\star > 10^{10}\, \mathrm{M}_\odot$, where black holes are massive enough to influence their host galaxies \citep{Nelson2019, Lapiner2024}.


The galaxy sample consists of 169 galaxies tracked in the redshift interval $5.25<z<9$, and whose haloes have a maximum circular velocity range between 50 and $300\, \mathrm{km \, s^{-1}}$. This covers a halo mass range between $10^9$ and a few times $10^{11} \, \mathrm{M_\odot}$, and more than 4 orders of magnitude in stellar mass, $10^6 < M^\star/\mathrm{M}_\odot < 10^{10.5}$. We store a total of 45 snapshots for each galaxy, with a time spacing of $7 - 10\,\mathrm{Myr}.$\footnote{The time step in the simulation is much shorter than these values, being typically $1000\,\mathrm{yr}$.}

The target haloes are initially selected using low-resolution \textit{N}-body-only simulations of three cosmological boxes with sizes 10, 20 and 40 $\rm \mathit{h}^{-1}Mpc$, assuming WMAP5 cosmology with matter density, baryon density, Hubble, and density fluctuation parameters $\Omega_\mathrm{m} = 0.27$, $\Omega_\mathrm{b} = 0.045$, $h = 0.7$, and $\sigma_8=0.82$, respectively \citep{Komatsu2009}. Initial conditions for the selected haloes are then generated using a standard zoom-in technique \citep{Klypin2011}, with a much higher resolution. The resulting resolution of dark matter particles is $m_\mathrm{DM} = 10^4 \, \mathrm{M}_\odot$, with a minimum stellar particle mass of  $100 \, \mathrm{M}_\odot$. The maximum spatial resolution is always between 8.7 and 17 proper parsecs (a comoving resolution of $109\,\mathrm{pc}$ after $z=11$). 

\section{Results}
\label{results}

\begin{figure}
\centering
\resizebox{\hsize}{!}{\includegraphics{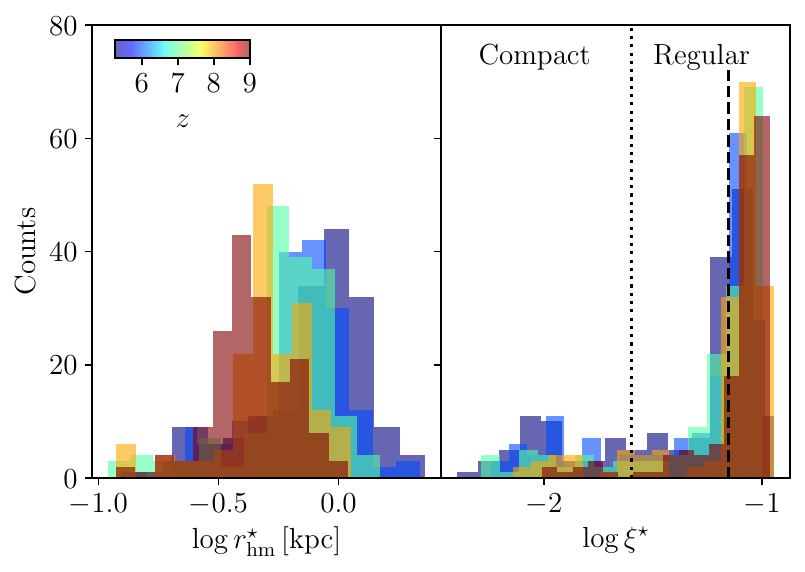}}
\caption{Distribution of the half-mass stellar radius (left panel) and the stellar-to-halo size ratio (right panel), for $5.25 < z < 9$ (colour coded). The dashed vertical line indicates the median value of $\xi^\star$  over the full sample and redshift range ($ \left< \xi^\star  \right> = 0.07$). A growing subpopulation of compact galaxies is seen; the dotted vertical line at $\log \xi^\star = -1.6$ roughly illustrates the separation.}
\label{radiusdistro}
\end{figure}

To represent the size of FL simulated galaxies, we use the half-mass radius of the distribution of stellar particles $r^\star_\mathrm{hm}$. This is defined as the radius of a sphere that encloses half of the stellar mass found within $0.15 r_{200}$, with $r_{200}$ the usual definition of the virial radius (i.e., the radius of a spherical volume with a mean overdensity of 200). We adopt this quantity because it is less influenced by geometric effects than 2D-projected sizes, and also by tracer biases like those affecting the sizes computed from the light in specific bands. For all galaxies we find at least 1000 stellar particles within $r^\star_\mathrm{hm}$ at any redshift, which makes our analysis numerically robust and allows us to achieve a well-sampled stellar mass distribution.

In Fig. \ref{radiusdistro} we show the distribution of $r^\star_\mathrm{hm}$. Its peak radius and width evolve to larger values with decreasing redshift, displaying the hierarchical growth of galaxies as they are assembled. To subtract the bulk of this effect from the general size evolution, we also present the distribution of the stellar-to-halo size ratio $\xi^\star = r^\star_\mathrm{hm} / r_{200}$. A bimodality is already apparent; most galaxies belong to the peak at $\log \xi^\star \approx -1$ (hereafter \textit{regular galaxies}). This subpopulation shows a very weak evolution with redshift, in the sense that haloes grow faster than the distribution of stars, which is expected because the latter is regulated by feedback whereas the former are not. The second peak at $\log \xi^\star \lesssim -1.6$ indicates the existence of a subpopulation of \textit{compact galaxies}. This peak grows as redshift decreases, indicating a continued production of these objects. The fact that our sample comprises only those galaxies that can be followed through the whole redshift interval implies that the compact subpopulation grows at the expense of the regular one. In other words, there are galaxies that become more compact as time elapses. The median value of $\xi^\star$ over the full redshift range ($5.25 < z < 9$) is $\left< \xi^\star \right> = 0.07$, similar to the value of 0.08 reported by \citet{Ma2018} for \textsc{fire} zoom-in simulated galaxies at comparable epochs ($6 < z < 10$). However, \citet{Ma2018} do not find evidence of bimodality in their sample. This is likely due to the fact that the majority of their galaxies are low-mass systems with $M^\star < 10^8\, \mathrm{M}_\odot$. Furthermore, they report an overall expansion in their galaxy population, in contrast with our findings.

\begin{figure}
\centering
\resizebox{9cm}{!}{\includegraphics{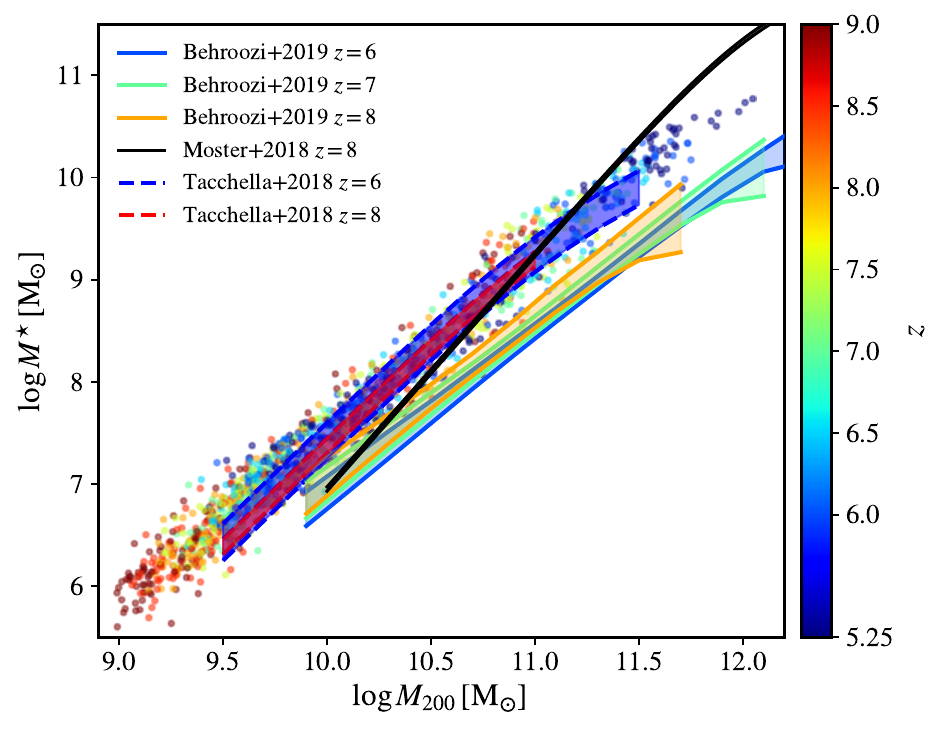}}
\resizebox{9cm}{!}{\includegraphics{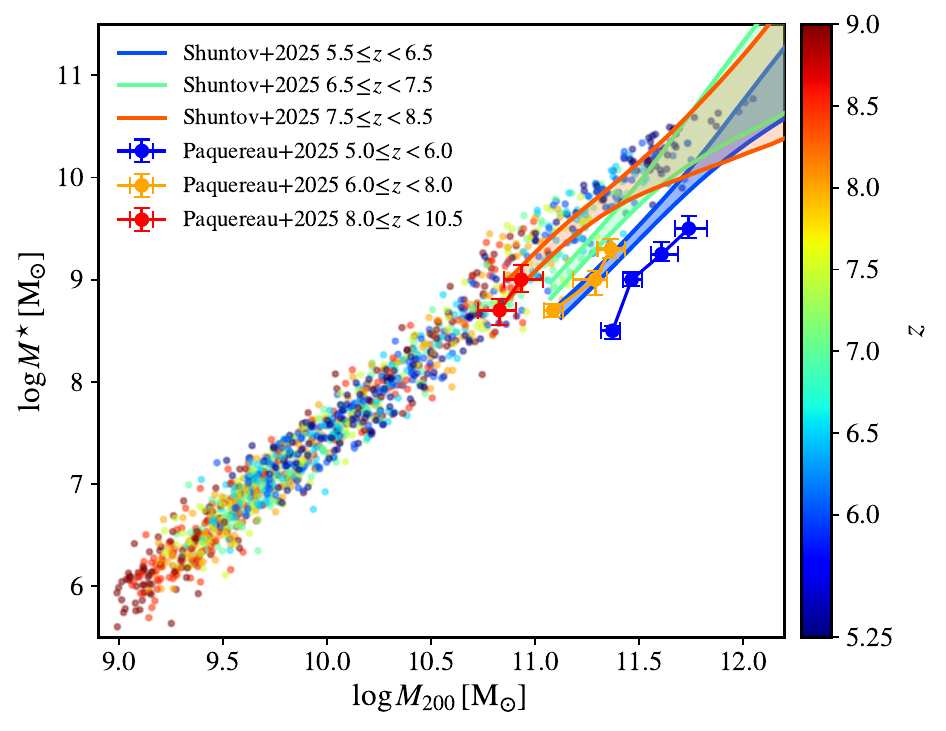}}
\caption{Relation between the stellar mass of galaxies and the virial mass of their haloes, compared to those predicted by semi-empirical models (top panel) and the Cosmos-Web survey subproducts (bottom panel). The shaded regions represent the dispersion of the models.}
\label{stellarhalomass}%
\end{figure}

Fig.~\ref{stellarhalomass} plots the relation between the stellar mass $M^\star$ of galaxies and the virial mass $M_{200}$ of their haloes. We define the former as the stellar mass within a radius enclosing 83\% of the baryonic mass of the galaxy (called the optical radius) and the latter as the total mass within $r_{200}$. FL galaxies follow a tight relation in the $M_{200}$--$M^\star$ plane, with a scatter of $\sim 1\, \mathrm{dex}$, assumed to be driven by the different mass accretion histories \citep{Moster2018}. On the one hand (top panel) this relation is consistent with the results of semi-empirical models of \citet{Tacchella2018} and \citet{Behroozi2019}, although with a small offset in the case of the latter due to the different high-$z$ samples used. It shows, however a small discrepancy at $M_{200}  \gtrsim  10^{11.5}\,\mathrm{M}_{\odot}$ (a tiny part of the FL sample at $z=5.25$) with the models of \citet{Moster2018}. Some modern cosmological simulations with phenomenological feedback models \citep{Pillepich2018,Dave2019} assume a decrease of the mass of galactic outflows with increasing galaxy mass to correct this discrepancy. However, these models are calibrated using observations at lower redshifts and may fail at earlier times, when gas densities were much higher. The FL suite, instead, predicts a higher density of massive galaxies ($M^\star >10^9 \, \mathrm{M}_{\odot}$) resulting in an improved agreement with JWST data in the range $z=6-13$, compared to the predictions of other models \citep{Ceverino2024}. On the other hand (bottom panel), the relation is in good agreement (except for an offset similar to that described above) with the results of the Cosmos-Web survey, obtained with both abundance-matching \citep{Shuntov2025} and HOD \citep{Paquereau2025} techniques. The tight relation observed in Fig.~\ref{stellarhalomass} implies that the FL virial masses, and therefore the virial radii, are typical of their stellar masses. We conclude that the compactness observed in a subpopulation of FL galaxies is due to a stellar distribution anomalously concentrated given its mass\footnote{As opposed to an anomalously extended halo.}. This sub-population must therefore arise from baryonic mechanisms.

\begin{figure}
\centering
\resizebox{9cm}{!}{\includegraphics{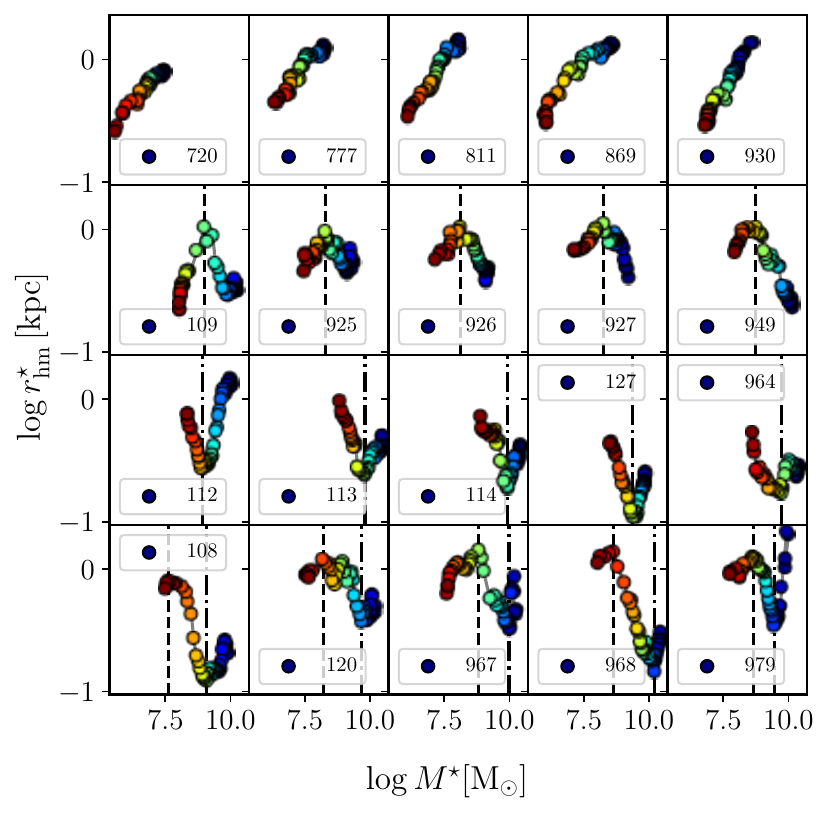}}
\caption{Different behaviours observed in the evolution of the sizes of individual galaxies: always expanding objects (top row), those reverting expansion into compaction (second row), systems reverting compaction into expansion (third row), and galaxies showing the complete ECE process (bottom row). Dashed and dot-dashed lines mark the masses at which galaxies attain their maximum and minimum sizes, respectively. Each panel represents a galaxy, labelled by its simulation ID.}
\label{subsampling}%
\end{figure}

To shed light onto the origin of the compact subpopulation, we explored the evolution of the sizes of the 169 individual galaxies in our catalogue. Fig.~\ref{subsampling} shows the four types of behaviour found: a) galaxies that always expand, b) objects showing an expansion at early times followed by a compaction at late times, c) systems compacting first and expanding later, and d) a combination of b) and c). We interpret this variety of behaviours as different stages of a single process in which galaxies first expand, then contract, and finally re-expand (hereafter \textit{ECE process}). The compact subpopulation comprises then those galaxies going through the end of the contraction stage and the beginning of the following re-expansion.

\begin{figure}
\centering
\resizebox{9cm}{!}{\includegraphics{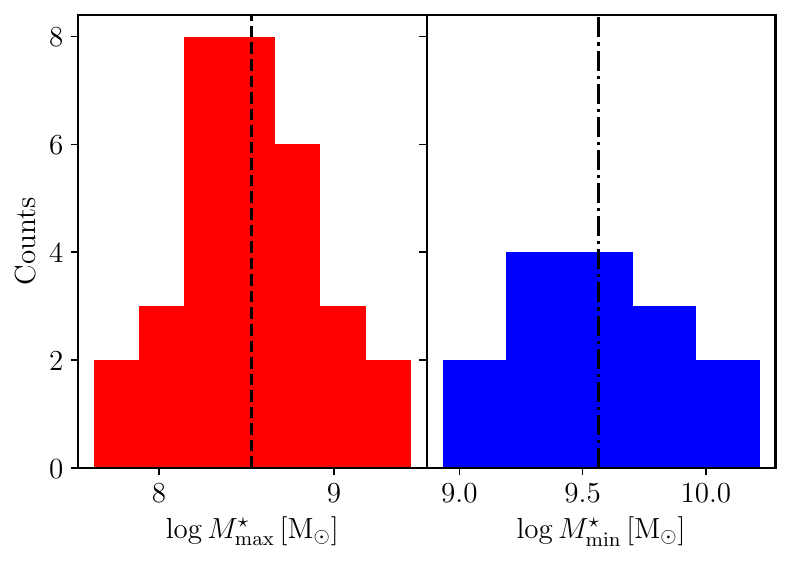}}
\caption{Distribution of the masses at which galaxies attain their peak (left panel) and minimum (right panel) sizes. The dashed and dot-dashed lines represent the geometric mean values $M^\star_\mathrm{on}$ and $M^\star_\mathrm{off}$, respectively.}
\label{turnondistro}%
\end{figure}

The distribution of the masses $M^\star_\mathrm{max}$ at which galaxies attain their peak sizes (Fig.~\ref{turnondistro}) reveals that compaction predominantly occurs around $ M^\star \sim 10^{8-9} \, \mathrm{M}_\odot$. We define the \textit{turn-on} mass $M^\star_\mathrm{on}$ as their geometric mean,

\begin{equation}
\log (M^\star_\mathrm{on} / \mathrm{M}_\odot) = \langle \log (M^\star_\mathrm{max} / \mathrm{M}_\odot) \rangle = 8.53. 
\end{equation}

\noindent
This represents the mass scale at which galaxies revert their initial expansion and begin to contract. The masses $M^\star_\mathrm{min}$ at which galaxies begin re-expansion are concentrated around $ M^\star \sim 10^{9-10} \, \mathrm{M}_\odot$  (Fig.~\ref{turnondistro}). We define the mean of this distribution as the \textit{turn-off} mass.

\begin{equation}
\log (M^\star_\mathrm{off} / \mathrm{M}_\odot) = \langle \log (M^\star_\mathrm{min} / \mathrm{M}_\odot) \rangle = 9.57.
\end{equation}

\noindent
We remark that the different phases of the ECE process occur at specific mass scales, indicating that galaxy stellar mass is its main driver. 

To better analyze the development of this process, we divided our sample of all 7605 snapshots into four subsamples (SS):

\begin{itemize}
\item SS1: includes galaxies that always expand. These serve as a reference sample, representing systems not affected by the process within the redshift interval studied.

\item SS2: comprises galaxies during the initial expansion phase, prior to compaction.

\item SS3: includes galaxies undergoing the compaction phase, characterized by size reduction.

\item SS4: consists of galaxies in the re-expansion phase, following the compaction event.
\end{itemize}

Note that, by construction, all galaxies except those in SS1 contribute to more than one SS; we include in each SS only the redshifts at which the galaxy displays the corresponding behavior. 

\begin{figure}
\centering
\resizebox{9cm}{!}{\includegraphics{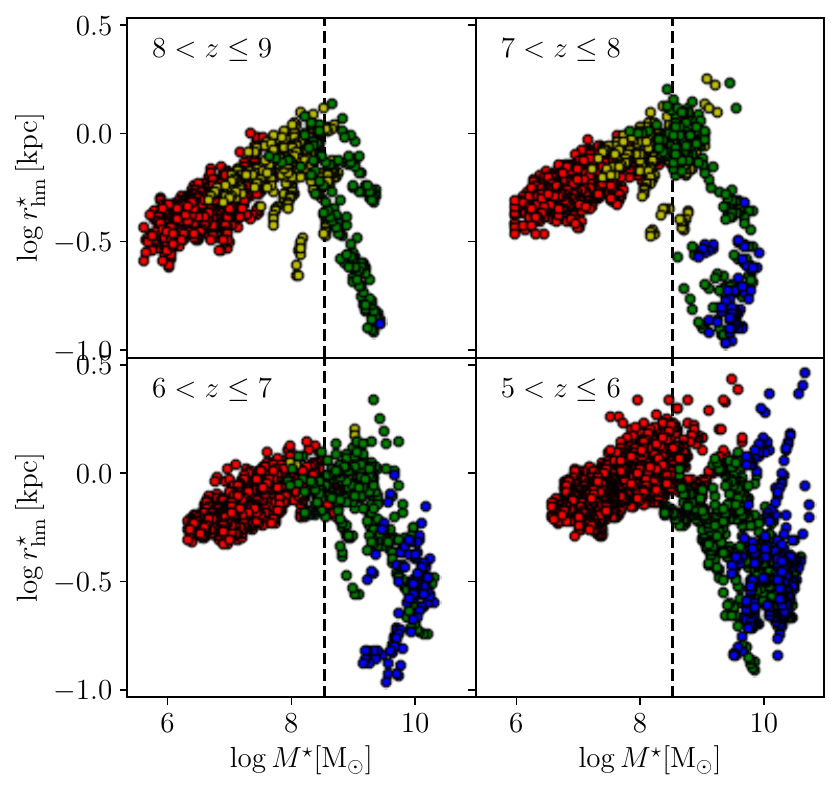}}
\caption{Dependence of the stellar size on stellar mass for different redshifts and galaxy subsamples (indicated with different colours, SS1: red, SS2: yellow, SS3: green, SS4: blue). The dashed vertical line indicates the value of the turn-on stellar mass, $M^\star_\mathrm{on} = 10^{8.53}\,\mathrm{M}_\odot$.}
\label{masssize}%
\end{figure}

\begin{figure}
\centering
\resizebox{9cm}{!}{\includegraphics{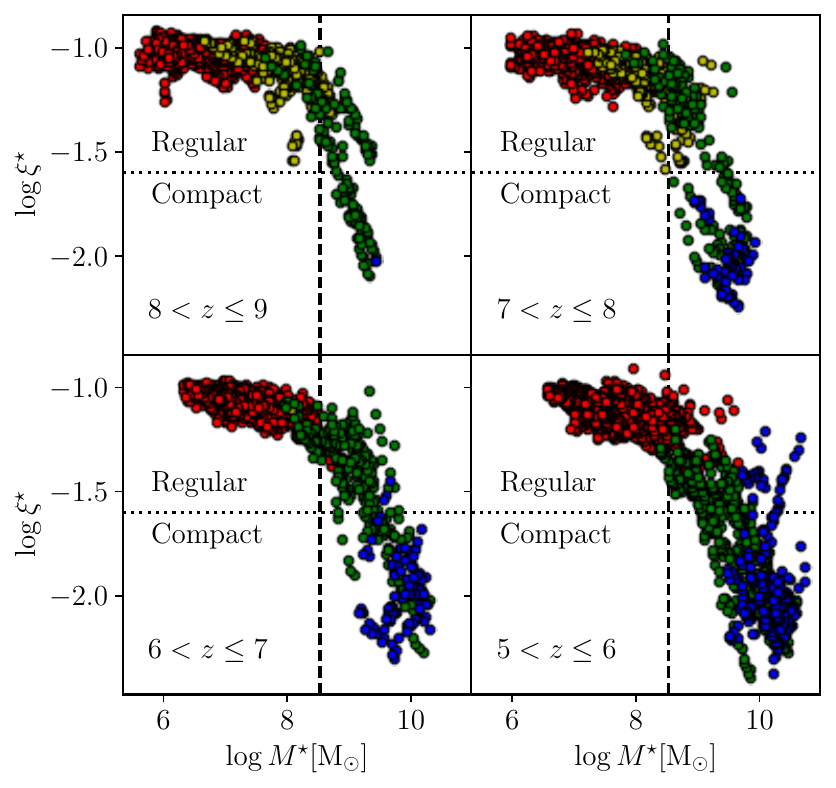}}
\caption{Dependence of stellar-to-halo size ratio on stellar mass for different redshifts and galaxy subsamples. References are as in Fig.~\ref{masssize}. The dotted horizontal lines separate regular from compact galaxies.}
\label{masscompactness}%
\end{figure}

Fig.~\ref{masssize} shows the stellar size--mass relation of our sample, for redshifts $z \in [5.25,9]$. Two different behaviours are already apparent at all redshifts: a tight ascending branch in which galaxy sizes increase with their masses \citep[hereafter \textit{normal} branch, because it follows the same trend observed at low redshifts,][]{Shen2003,vanderWel2014,Mowla2019}, and a descending one with the opposite behaviour and a dispersion that increases with decreasing redshift. The break separating both branches occurs at $M^\star \approx M^\star_\mathrm{on}$, suggesting that the change in the trend is connected to the beginning of the compaction phase of galaxies. Indeed, galaxies in SS1 and SS2 generally lie at masses lower than $M^\star_\mathrm{on}$, whereas the opposite is true for SS3 and SS4, clearly showing once again that this mass scale is a global proxy for the onset of compaction. Above $M^\star_\mathrm{on}$, only $18\%$ of galaxies are expanding, whereas $58\%$ are contracting and the remaining $24\%$ have already contracted and are experiencing re-expansion. A second ascending branch due to the re-expansion of galaxies is not observed because the scatter in $M^\star_\mathrm{min}$ mixes compacting and re-expanding galaxies at a given mass, producing a dispersion in the descending branch instead. A wider redshift range, allowing the exploration of the whole re-expansion phase, would be needed to determine if a second ascending branch does exist.

These trends are clear also in the relation between the stellar-to-halo size ratio $\xi^\star$ and the stellar mass (Fig.~\ref{masscompactness}). This relation has a negative slope at all masses, originated in the already discussed faster growth of haloes with respect to the stellar distribution. A break separates the low- and high-mass branches at $M^\star_\mathrm{on}$, the latter being steeper than the former. All galaxies with masses below \( M^\star_\mathrm{on} \) belong to the regular subpopulation ($\log \xi^\star > -1.6$); compact galaxies are found only in the high-mass branch.

\begin{figure}
\centering
\resizebox{9cm}{!}{\includegraphics{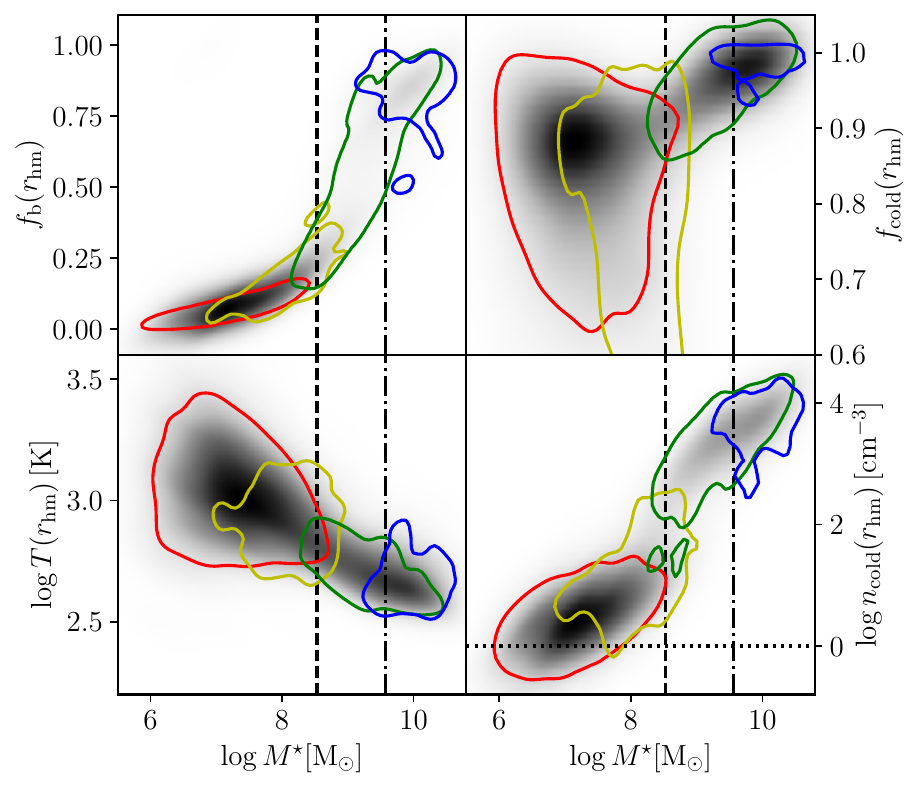}}
\caption{Dependence of different properties of baryons on the stellar mass. Top panels: the fraction of baryons (left) and cold gas (right) within $r^\star_\mathrm{hm}$. Bottom panels: density-weighted mean gas temperature (left) and mean density (right) within $r^\star_\mathrm{hm}$. The gray shades represent the density of galaxies in the plane defined by stellar mass and the corresponding property, computed from the data using a Gaussian smoothing kernel (normalization is different for each panel). Colour contours enclose $80\%$ of the systems in different subsamples (SS1: red, SS2: yellow, SS3: green, SS4: blue). The dashed and dash-dotted vertical lines mark $M^\star_\mathrm{on}$ and $M^\star_\mathrm{off}$, respectively, whereas the dotted horizontal line represent the density threshold for star formation.}
\label{sizemassvsproperties}%
\end{figure}

\begin{figure}
\centering
\resizebox{9cm}{!}{\includegraphics{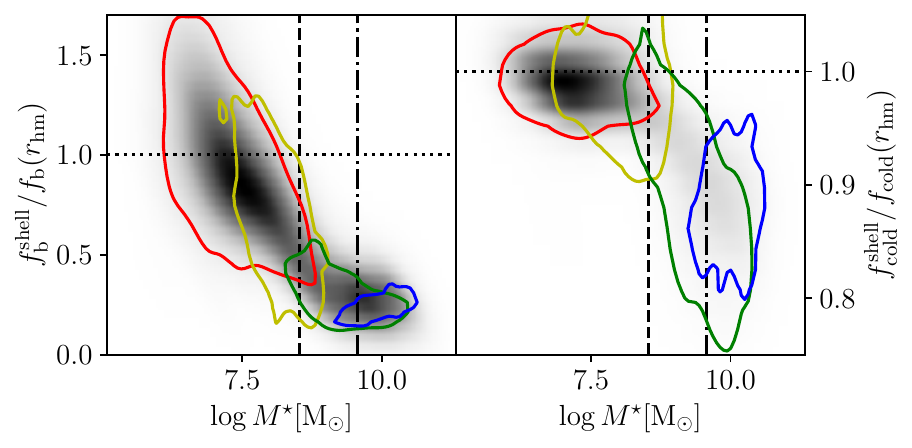}}
\caption{Ratio of the fraction of baryons (left panel) and cold gas (right panel) in a shell between $r^\star_\mathrm{hm}$ and $0.15\, r_{200}$ to those within $r^\star_\mathrm{hm}$, as a function of stellar mass. References are as in Fig.~\ref{sizemassvsproperties}. Dotted horizontal lines mark unit ratios.}
\label{baryonsoutside}%
\end{figure}

In order to explore the physical nature of the compaction mechanism and what conditions trigger it, in Fig.~\ref{sizemassvsproperties} we plot the dependence of different baryon properties on stellar mass. The fraction of baryons and cold gas (i.e., gas at temperature $T < 10^4 \,\mathrm{K}$) within radius $r$ are defined as $f_\mathrm{b}(r) = M_\mathrm{b}(r) / M_\mathrm{t}(r)$ and $f_\mathrm{cold}(r) = M_\mathrm{cold}(r) / M_\mathrm{g}(r)$, respectively. Here $M_\mathrm{b}(r)$, $M_\mathrm{t}(r)$, $M_\mathrm{g}(r)$, and $M_\mathrm{cold}(r)$ are respectively the total, baryonic, gas, and cold gas masses enclosed inside $r$. The mean temperature of cold gas $T_\mathrm{cold}(r)$ is computed as the density-weighted average within the same volume, whereas its mean density is $n_\mathrm{cold}(r) = 3 M_\mathrm{cold}(r) / 4 \pi r^3 m_\mathrm{H}$, with $m_\mathrm{H}$ the mass of the hydrogen atom. 

When evaluating these properties within $r^\star_\mathrm{hm},$ there are clear differences between galaxies of different subsamples. The sequence SS1--SS2--SS3--SS4 shows progressively higher baryon fractions from $f_\mathrm{b}(r^\star_\mathrm{hm}) \sim 0.25$ up to almost unity (top left panel of Fig.~\ref{sizemassvsproperties}), with a large jump at the turn-on mass scale. This suggests that the ECE process is related to the transition of the central regions of galaxies from dark-matter-dominated to baryon-dominated systems. It might be argued that the change in the baryon fraction could be interpreted as a consequence of the inward displacement of $r^\star_\mathrm{hm}$, because central regions contain more baryons. However, an increase in $f_\mathrm{b}(r^\star_\mathrm{hm})$ with $M^\star$ is already seen in the reference SS1, that comprises galaxies that never contract. 

Our results point therefore to the existence of an infalling flow of gas reaching the central regions, that modifies the ratio of baryons to dark matter, in agreement with previous works \citep{Zolotov2015}. Under this interpretation (for which we will provide further evidence below), Fig.~\ref{sizemassvsproperties} (top right panel) shows also that the fraction of the cold gas phase increases from $f_\mathrm{cold}(r^\star_\mathrm{hm}) \sim 0.9$ up to $\sim 1$. This happens because the infalling flow comprises mainly cold gas and/or the enhanced density accelerates cooling. Moreover, during compaction (SS3) and the following re-expansion (SS4), almost all gas in the central regions of galaxies is cold. This gas has reached conditions for star formation (temperature $T < 10^4\, \mathrm{K}$, density $n_\mathrm{cold} > 1\,\mathrm{cm}^{-3}$, bottom panels of Fig.~\ref{sizemassvsproperties}) in almost all systems except for the less massive ones in SS1. Indeed, this medium becomes cooler and denser as the ECE process develops, which reduces the local free-fall time and thus shortens the star formation timescale. As a result, SFR density is strongly enhanced.

Our results point therefore to a wet compaction scenario \citep{Dekel2014,Zolotov2015}, characterized by the infall of large amounts of cold gas from the outer regions of galaxies, which compresses and cools up to the point that a starburst develops in the innermost regions. If the starburst is strong enough and occurs near the galaxy centre, it may reverse the normal inside-out star formation, turning the expansion of the stellar distribution into compaction. However, in previous works compaction typically occurs at lower redshifts ($z \sim 2-4$) and a higher critical stellar mass \citep[$\sim 10^{10}\, \mathrm{M}_\odot$;][]{Dekel2014,Zolotov2015,Tacchella2016}. While driven by the same underlying mechanisms (self-reinforced cold gas inflows and central starburst activity) the ECE process identified in our simulations takes place earlier in the history of the Universe ($9 > z > 5$) and at a lower mass scale. It may indeed be responsible for the formation of spheroidal structures that would become the bulges of low-redshift galaxies \citep[cf.][]{Roper2022}. After compaction, conditions may again favour star formation over a more extended region, resuming the inside-out growth of galaxies. Given that our ECE process requires that the stellar mass in the central regions exceeds the dark mass, and that the former is directly related to the total stellar mass, it is hardly conceivable that the differences with previous works, especially the critical masses, are due to different physics or calibration issues. Most probably we are seeing different stages in the evolution of the galaxies at which the wet compaction mechanism plays an important role.

Three issues must be explored to underpin the scenario described above. First, the time and spatial evolution of the process must agree with the described scenario. Second, the reason behind the efficient gas cooling that makes the starburst massive must be found. Third, the mechanism driving the strong gas infall must be pinpointed. Fig.~\ref{baryonsoutside} shows the fraction of baryons $f^\mathrm{shell}_\mathrm{b}$ and cold gas $f^\mathrm{shell}_\mathrm{cold}$ in a shell with inner and outer radii $r^\star_\mathrm{hm}$ and $0.15\, r_{200}$ respectively, compared to the same fractions within $r^\star_\mathrm{hm}$. The outer shell radius is well beyond $r^\star_\mathrm{hm} \lesssim 0.1 \, r_\mathrm{200}$ for all galaxies (Fig.~\ref{radiusdistro}) and roughly marks the outer boundary of the stellar distribution. The ratios are near unity in SS1, demonstrating the existence of similar conditions in the central and outer regions of galaxies. They progressively drop by a significant amount in SS2--SS3--SS4, indicating that the key changes in baryonic properties responsible for driving the ECE process take place exclusively in the central regions.

\begin{figure}
\centering
\resizebox{9cm}{!}{\includegraphics{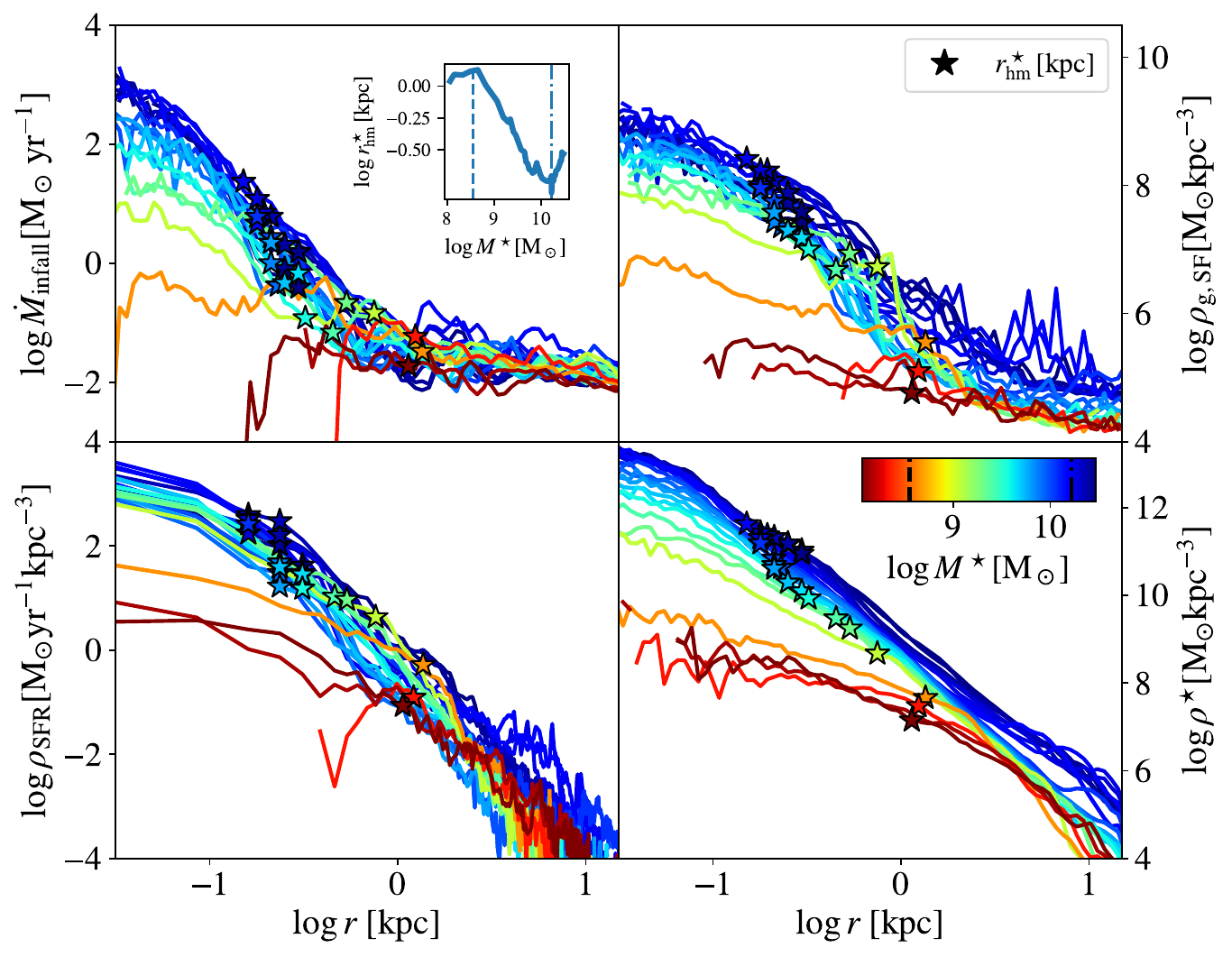}}
\caption{Radial profiles of the infalling mass flux (top left panel), and the densities of star-forming gas (top right panel), star formation rate (bottom left panel), and stars (bottom right panel), for an individual galaxy across its evolution. Colours refer to the evolving stellar mass of the galaxy. The inset shows the evolution of the galaxy size.}
\label{radialprofiles}%
\end{figure}

The picture described above is supported also by the analysis of the evolution of the radial distribution of baryons in individual galaxies. Fig.~\ref{radialprofiles} (top left panel) shows the radial flux of cold gas $\dot{M}_\mathrm{infall}$ (defined as positive when gas flows inwards) as a function of radius and stellar mass, for a typical galaxy during the SS2--SS3--SS4 phases. It confirms the development of a strong infalling flow when the system reaches a stellar mass of the order of $M^\star_\mathrm{on}$. The flux is significant only well within $r^\star_\mathrm{hm}$ during the whole evolution, thus discarding that the changes we are seeing in baryon properties are merely the consequence of looking closer to the centre as $r^\star_\mathrm{hm}$ shrinks. This flow carries cold gas to the inner kiloparsec, increasing the density of star-forming gas  ($\rho_\mathrm{g,SF}$; Fig.~\ref{radialprofiles}, top right panel), and thus triggering a localised star formation episode that leads to a central cusp in the SFR density profile ($\rho_\mathrm{SFR}$; Fig.~\ref{radialprofiles}, bottom left panel). The gas densities ensure a high SFR via the Schmidt-Kennicutt law, making newborn stars dominate the stellar population over old ones. This produces a more compact stellar density distribution ($\rho^\star$; Fig.~\ref{radialprofiles}, bottom right panel) and therefore a lower $r^\star_\mathrm{hm}$.

\begin{figure}
\centering
\resizebox{9cm}{!}{\includegraphics{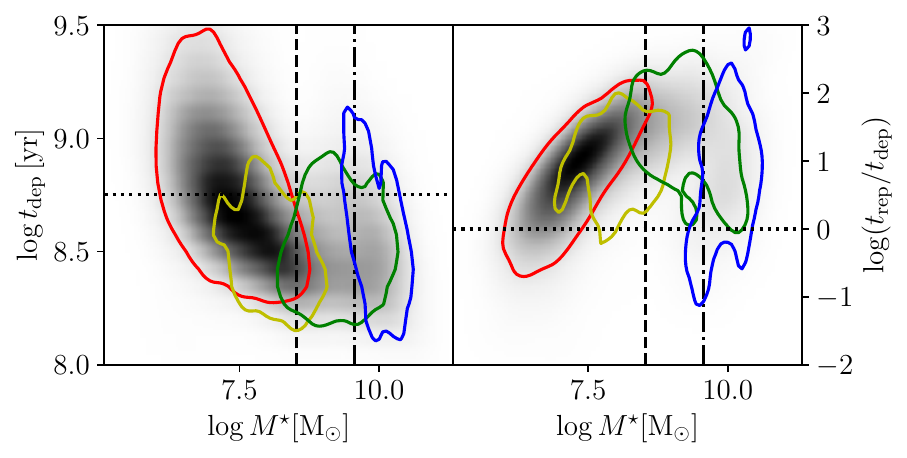}}
\caption{Depletion time (left panel) and the ratio of replenishment-to-depletion times (right panel) within $r^\star_\mathrm{hm}$, as a function of stellar mass. References are as in Fig.~\ref{sizemassvsproperties}. The dotted horizontal line in the left panel shows the timespan of the simulations, whereas that in the right panel marks the unit ratio.}
\label{depletion}%
\end{figure}

The aforementioned scenario explains the compaction phase of the process as originated in a self-reinforced infall episode that drives a sudden change of the prevailing gas conditions in the inner regions of galaxies. The recovery of an expansion phase, instead, is not led by a subsequent reverse change in these conditions. Galaxies in SS4 follow the same trends of those in SS2--SS3, and indeed represent extreme values with respect to the latter (Fig.~\ref{sizemassvsproperties}). The profiles in Fig.~\ref{radialprofiles} show not only that the infall persists at late times, but also that the flux reaches higher values. The cold gas, SFR and stellar densities continue to increase at the centre, but develop growing tails up to larger radii. This suggests that the expansion is due to the enlargement of the region in which cold gas reaches star formation conditions.
This re-expansion is different from that proposed by previous works \citep{Dekel2014,Zolotov2015,Tacchella2016}. Although these authors find a similar mass scale for re-expansion, they argue that this phase is driven by central gas depletion, that leads to the quenching of star formation in the core. Stars are then formed in a ring fuelled by fresh gas accreted at larger radii, resuming the inside-out growth of the galaxy. The fact that our galaxies reach the critical mass scale at earlier times suggests that they are richer in gas, which makes depletion less probable.

Fig.~\ref{depletion} shows the evolution of the gas depletion and replenishment times in the central regions, $t_\mathrm{dep} = M_\mathrm{g}(r^\star_\mathrm{hm}) / \dot{M}_\mathrm{SFR} (r^\star_\mathrm{hm})$ and $t_\mathrm{rep} = M_\mathrm{g}(r^\star_\mathrm{hm}) / \dot{M}_\mathrm{infall}(r^\star_\mathrm{hm})$ respectively, where $\dot{M}_\mathrm{SFR} (r)$ is the total star formation rate within radius $r$. Expanding galaxies (SS1--SS2) display progressively lower depletion times due to the decrease of their gas reservoir, in line with the findings of \citet{Ceverino2018}. However, $t_\mathrm{dep}$ falls barely below the timespan of the simulations at the end of this stage, indicating that this reservoir is never completely exhausted. The decrease of $t_\mathrm{dep}$ stops as soon as compaction starts, driven by the fresh gas infall. During compaction and re-expansion stages (SS3--SS4) the depletion time remains almost constant, whereas the replenishment time decreases as infall becomes increasingly relevant. Both times become almost equal at the latest stage (SS4), indicating that the gas in the central region is replaced by fresh material from outside at a similar rate that it is used in star formation, and therefore never completely depleted. The existence of outflows may enhance depletion but would hardly modify this result, as outflow rates are usually similar to the SFR \citep{Zolotov2015}.

\begin{figure}
\centering
\resizebox{9cm}{!}{\includegraphics{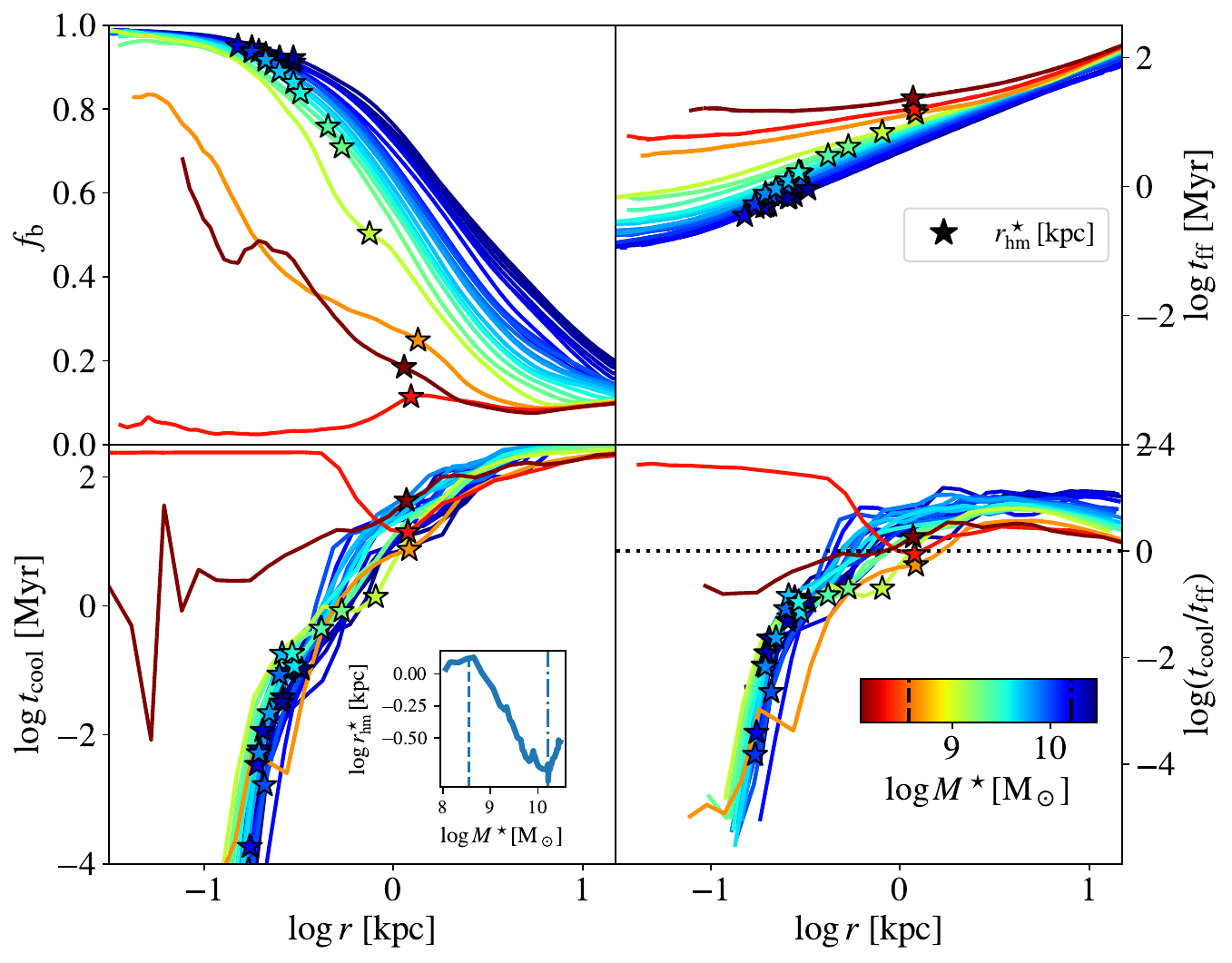}}
\caption{Radial profiles of the fraction of baryons (top left panel), the free-fall (top right panel) and cooling times (bottom left panel), and the ratio $t_\mathrm{cool}/t_\mathrm{ff}$ (bottom right panel), for the same galaxy of Fig.\ref{radialprofiles}. The dashed horizontal line marks $t_\mathrm{cool}=t_\mathrm{ff}$.}
\label{radialprofiles2}%
\end{figure}

Fig.~\ref{radialprofiles2} provides clear evidence regarding the conditions that make possible the strong starburst found in our previous results. For the same typical galaxy of Fig.~\ref{radialprofiles}, the increase in the mass of the central regions driven by the infall of baryons (Fig.~\ref{radialprofiles2}, top left panel) causes a decrease of the free-fall time $t_\mathrm{ff}$ in the same region (Fig.~\ref{radialprofiles2}, top right panel). This enhances the infall itself, creating a self-reinforced flow of gas that increases the density in the central regions by several orders of magnitude. This gas cools in a time scale

\begin{equation}
t_\mathrm{cool} = \frac{3.45 \, k_\mathrm{B} \, T}{n_\mathrm{g} \, \Lambda(n_\mathrm{g}, T, Z)},
\label{eq:tcool}
\end{equation}

\noindent where $k_\mathrm{B}$ is Boltzmann constant, $n_\mathrm{g}$ and $Z$ are the number density and metallicity of the gas, respectively, and $\Lambda$ is the cooling rate per unit density. The cooling rate shows a strong dependence of $n_\mathrm{g}$ \citep{Wiersma2009}, ensuring that in the central regions the cooling time is short (Fig.~\ref{radialprofiles2}, bottom panels) and almost all gas remains cold. The increase in density produced by the infall thus causes a massive starburst that makes the stellar distribution more compact.

\begin{figure}
\centering
\resizebox{9cm}{!}{\includegraphics{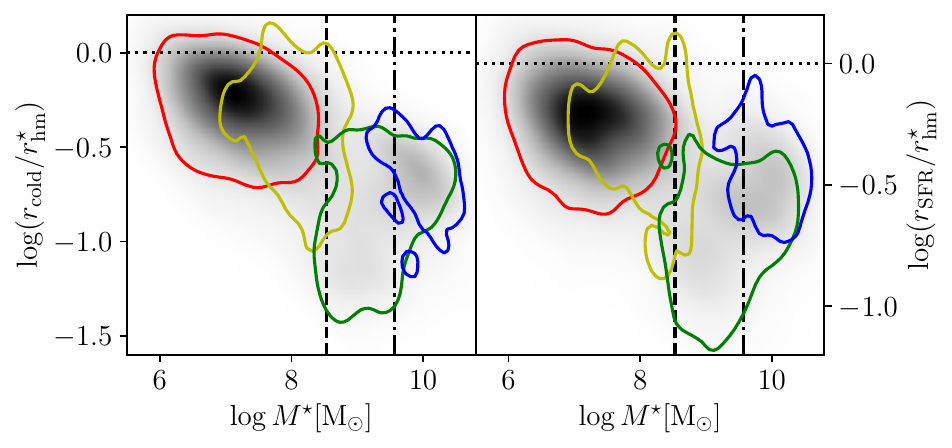}}
\caption{Dependence on the stellar mass of the density-weighted average radii of the distribution of cold gas (top left panel) and SFR (top right panel). References are as in Fig.~\ref{sizemassvsproperties}.}
\label{meanradii}%
\end{figure}

Fig.~\ref{meanradii} provides further evidence for our scenario by displaying the dependence on stellar mass of the density-weighted average radii of the cold gas and SFR distributions ($r_\mathrm{cold}$ and $r_\mathrm{SFR}$, right and left panels, respectively). These radii estimate the size of the regions where the cold gas lies and the stars are being formed, respectively. They are computed as

\begin{equation}
r_\mathrm{cold}= \left< r \right>_{\rho _{\mathrm{cold}}}= \frac{\int_{0}^{0.15\, r_{200}} r \rho _{\mathrm{cold}} (r) \, dr}{\int_{0}^{0.15\, r_{200}}\rho _{\mathrm{cold}} (r) \, dr},
\end{equation}

\noindent
and

\begin{equation}
r_\mathrm{SFR}= \left< r \right>_{\rho _{\mathrm{SFR}}} = \frac{\int_{0}^{0.15\, r_{200}} r \rho _{\mathrm{SFR}} (r) \, dr}{\int_{0}^{0.15\, r_{200}}\rho _{\mathrm{SFR}} (r) \, dr}.
 \end{equation}

\noindent The upper limit of integration is set to $0.15\, r_{200}$, which approximately corresponds to the outer boundary of the stellar distribution. 
Both $r_\mathrm{cold}$ and $r_\mathrm{SFR}$ are close to $r^\star_\mathrm{hm}$ for galaxies in SS1, indicating that the star formation is widely distributed, which is consistent with the slow expansion of these galaxies. These radii fall well below $r^\star_\mathrm{hm}$ in SS2 and SS3, indicating that the star formation concentrates well inside $r^\star_\mathrm{hm}$, as proposed. Moreover, galaxies in SS2 are still expanding; the decrease in the half-mass radius lags behind those of the other radii, which is consistent with the latter decrease being the cause of the former. Fig.~\ref{meanradii} also shows that the cold gas and SFR distributions of SS4 galaxies are more extended with respect to the half-mass radius than those of SS3, indicating that star formation conditions proceed in SS4 into a broader region, which causes the recovery of an expansion phase.

\begin{figure}
\centering
\resizebox{9cm}{!}{\includegraphics{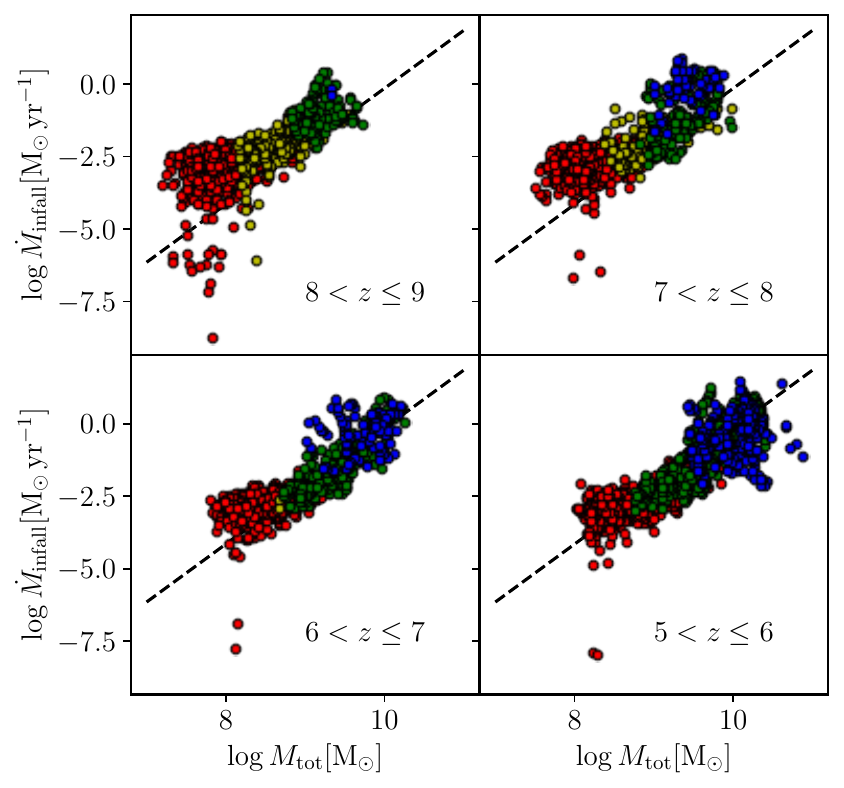}}
\caption{Flux $\dot{M}_\mathrm{infall}$ of infalling gas at $r^\star_\mathrm{hm}$ as a function of total mass $M_\mathrm{tot}$ within this radius, for different redshift intervals and galaxy subsamples (indicated with different colours, SS1: red, SS2: yellow, SS3: green, SS4: blue). Dashed lines represent a semiempirical model for the data, describing the infalling flux of gas in the Bondi--Hoyle regime.}
\label{infall}%
\end{figure}

Finally, the mechanism behind the self-reinforced infall--cooling--star formation episode can be unveiled from Fig.~\ref{infall}. The infalling flux $\dot{M}_\mathrm{infall}$ at $r^\star_\mathrm{hm}$ is roughly dependent on the square of the total mass $M_\mathrm{tot}$ within this radius, which is reminiscent of Bondi--Hoyle accretion \citep{Bondi1944}. Indeed, a simple model (dashed lines in Fig.~\ref{infall}) constructed by assuming that the flow proceeds in this regime provides a reasonably good agreement with the data at all redshifts. In this model,

\begin{equation}
\dot{M}_\mathrm{infall} = \frac{\pi \rho (1+z)^3 G^2 M_\mathrm{tot}^2}{(k_\mathrm{B} T m_\mathrm{H}^{-1})^{3/2}},
\end{equation}

\noindent
where $G$ is the gravitational constant and $m_\mathrm{H}$ is the mass of the hydrogen atom. We assume that gas reaches $r^\star_\mathrm{hm}$ from outside at a typical temperature $T = 10\,000\,\mathrm{K}$ and density $\rho = f_\mathrm{b,typ} \Omega_\mathrm{b} \rho_\mathrm{crit}$, where $\rho_\mathrm{crit,0} = 9.47 \times 10^{-27}\, \mathrm{kg\,m}^{-3}$ is the critical density of the Universe at redshift $z=0$, and $f_\mathrm{b,typ} = 0.05$ was taken as the mean baryon fraction of SS1 galaxies at $0.15\, r_{200}$. While the galaxies remain dark-matter dominated, the infalling flux depends mainly on the dark matter mass in the inner regions. As soon as baryons dominate, the flux couples to the baryonic mass provided by itself, and the dependence of Bondi--Hoyle accretion on $M_\mathrm{tot}^2$ generates the self-reinforced process observed. This change of behaviour happens roughly at the turn-on mass scale. 

\begin{figure}
\centering
\resizebox{9cm}{!}{\includegraphics{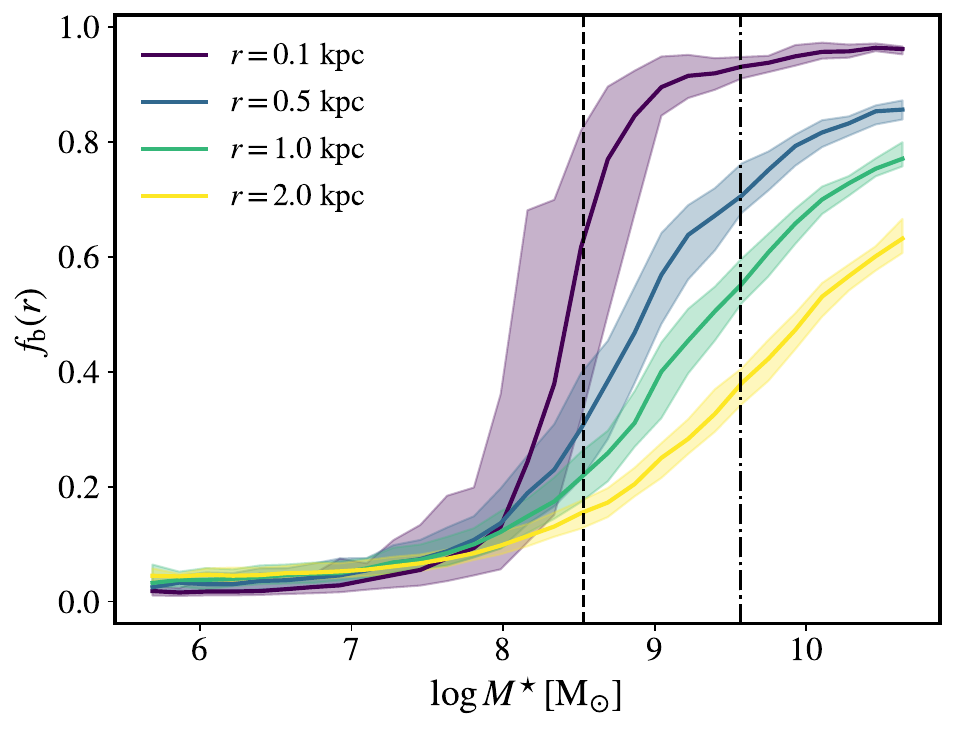}}
\caption{Fraction of baryons within spheres of different radii $r$ (indicated by different colours) as a function of stellar mass. Solid lines display the median over all galaxies, whereas shaded regions represent the dispersion of each distribution.}
\label{baryonfractionall}%
\end{figure}

This accretion mechanism sheds light also on the development of the re-expansion phase. Fig.~\ref{baryonfractionall} shows the change of the mean baryon fraction $f_\mathrm{b}(r)$ within spheres of progressively larger radii $r$. It is clear that the transition of central regions from dark-matter- to baryon-dominated proceeds in an inside-out way. The process starts near $M_\mathrm{on}$ only at the innermost regions, which increase their $f_\mathrm{b}$ and develop a sizeable inflow of gas that leads to a strong, concentrated starburst. This causes the stellar half-mass radius to shrink, as discussed above. As baryon dominance spreads outwards the self-reinforced accretion flow switches on at progressively larger radii, leading to a wider starburst and a less concentrated distribution of newly born stars. This stops and eventually reverses the decrease of $r^\star_\mathrm{hm}$, thus driving the observed re-expansion.

In summary, the three issues raised above have been settled. The time and spatial evolution of our simulated galaxies strongly support the wet compaction scenario, with a final re-expansion driven by the enlargement of the star-forming region instead of central gas depletion. Gas infall into the central regions of galaxies is the main driver of the process; the dependence of the infalling flux on the square of the mass of the accreting region is its key. At low stellar masses and baryon fractions, the flux depends mainly on the dominant dark-matter mass, and therefore proceeds at a slow rate. As soon as this slow infall begins to transform the central region of a galaxy into a baryon-dominated system, the dominant role is taken by the baryon mass. Infall becomes then faster through self-reinforcement, as it increases the same baryon mass that drives it. The result is a strong enhancement of the gas density in the central region, a large drop in the gas cooling time, and the consequent increase in the star formation rate. These produce a strong, concentrated starburst that makes the stellar distribution more compact.

\section{Discussion and conclusions}
\label{conclusions}

Recent JWST observations have challenged our understanding of galaxy formation during the Cosmic Dawn. Specifically, the physical mechanisms creating compact galaxies with sizes of the order of $100\, \mathrm{pc}$, and those leading to an inverted stellar mass--size relation at this epoch, are still a matter of debate. Aiming at unveiling those mechanisms, in this paper we have analysed the evolution of the stellar distribution of 169 galaxies from the \textsc{FirstLight} zoom-in simulation suite, which enables a statistically robust analysis of these early evolutionary processes.

We found that, underneath the bulk evolution due to hierarchical growth, a population of compact galaxies with sizes of the order of $\sim 100\,\mathrm{pc}$ develops. This population is a result of a process that stops the normal inside-out growth of galaxies, suddenly shrinking their stellar distribution to sizes comparable to those observed by JWST \citep{Baggen2023,Baker2024,Tacchella2024,Morishita2024}. Some of the observations are indeed compatible with our results in both size and mass \citep[e.g.][Fig.~4]{Morishita2024}. After this rapid compaction, the stellar mass distribution resumes its inside-out development. Remarkably, both the onset of compaction and the resumption of expansion occur near specific stellar mass scales, $M^\star_\mathrm{on} = 10^{8.53}\,\mathrm{M}_\odot$ and $M^\star_\mathrm{off} = 10^{9.57}\,\mathrm{M}_\odot$ respectively, suggesting that stellar mass is the main driver of the process. This happens because compaction begins when the stellar mass in the central regions, directly related to the total stellar mass, matches the dark mass in the same regions. Variations in the critical masses may therefore be due to different individual star formation histories and spatial distributions produced by, e.g., the environment or the merger history. Indeed, the mechanism operates in all galaxies that reach the corresponding mass scales during the analysed redshift interval. Galaxies in which we do not observe it have masses below $M^\star_\mathrm{on}$ at the lowest redshift, whereas galaxies with masses above $M^\star_\mathrm{off}$ at the highest redshift are all compact. This leads us to propose that the process is universal; which may be investigated further by enlarging our redshift interval and simulation box. We are working on this issue at present. 

The described process naturally leads to an inversion of the size--mass relation in the stellar mass range $10^{8.5} - 10^{9.5}\, \mathrm{M}_\odot$. This result is consistent with the observations of \citet{Baggen2023}, and with recent numerical works based on simulations such as \textsc{BlueTides} at $z = 7$ \citep{Roper2022,Marshall2022}. However, it is at odds with the conclusion of \citet{Morishita2024} who report a positive slope in the whole mass range, and also with other zoom-in simulations such as \textsc{Thesan-zoom} \citep{McClymont2025} and \textsc{FIREbox} \citep{Feldmann2023,Rohr2022}, whose results are consistent with such trend.
The inversion in the mass-size relation has been previously reported in numerical simulations by \citet{Zolotov2015} and \citet{Lapiner2024}. Both studies observed that, for certain mass ranges, the size exhibits a decrease with mass followed by the typical increase. In the case of \citet{Lapiner2024}, the mass ranges corresponding to this behaviour are similar to those found in our study; in other cases it is slightly higher ($10^{10}\, \mathrm{M}_\odot$). A notable difference, however, lies in the redshift coverage: \citet{Lapiner2024} explores the range from $z \approx 1$ to $5$, while \citet{Zolotov2015} focuses on $z \approx 2$ to $4$. In this regard, our work provides new insights by extending the analysis to higher redshifts, from $z \approx 5$ to $9$, with good resolution. This extended range allows us to capture the onset of the compaction process, and therefore the full expansion–-compaction–-re-expansion behaviour --- our sample is capable of tracing the entire cycle of size evolution. The agreement of the results throughout the whole interval $1 \lesssim z \lesssim 9$ supports our claim that this process is universal.

Regarding the physics, our results point to a wet compaction scenario similar to those described in previous works \citep{Dekel2014,Zolotov2015}. The infall of gas from the outer skirts of a galaxy to its innermost ($< 1\,\mathrm{kpc}$) region increases the density of the latter and changes its state from dark matter- to baryon-dominated. It also decreases the cooling time of the gas, enhancing the fraction of cold gas, and boosts star formation. The process becomes self-reinforced because the infalling flux increases with the baryon mass in the inner regions. The result is a strong central starburst, with newborn stars dominating the whole stellar distribution and making it to appear more compact.

Although the development of gas cooling and star formation conditions proceeds inside-out, the starburst is strong and concentrated enough to produce a rapid and steady decrease in the stellar half-mass radius. The subsequent re-expansion phase is driven by the broadening of the region in which baryons dominate, a natural consequence of gas flowing inwards, and the resulting enhancement of accretion at progressively larger radii. Therefore, cold gas can reach star formation conditions farther out from the centre, leading to a wider stellar distribution and a larger half-mass radius. This is a major difference with previous works, in which it is due to the depletion of gas and the corresponding quenching of star formation in the central regions, which leads to a star-forming ring that grows outward. We argue that the galaxies at low redshift, such as those studied in previous works, have low gas densities that facilitate quenching. Our high-redshift galaxies are gas-rich, which makes it difficult to exhaust their gas; it continues to form stars while more gas reaches the inner regions, expanding the star-forming region. Given that re-expansion is attained at the same mass scale but different redshifts, the development or quenched or normal re-expansion might depend on the environment. This issue deserves further exploration.

More recently, using the \textsc{Thesan} zoom-in simulations, \citet{McClymont2025} find that the evolution of galaxy sizes is closely linked to bursty star formation at high redshift. In their simulations, galaxies located above the star-forming main sequence tend to exhibit centrally concentrated star formation, which drives morphological compaction episodes. These central starbursts subsequently quench in an inside-out manner, resulting in cyclical alternations between compaction and expansion that cause galaxies to oscillate around an overall normal size–-mass relation, instead of a producing a steady compaction followed by an expansion. \citet{McClymont2025} claim that the reason behind the discrepancy between their results and those of other simulations is their greater mass resolution that produces a set of small starbursts instead of a single, massive one. It may be argued, however, that the overall increase in the stellar mass of galaxies is the same in all simulations, therefore the different behaviour of $r^\star_\mathrm{hm}$ is not due to mass resolution but to the varying spatial distribution of the starbursts with time. Fig. 4 of \citet{McClymont2025} shows starbursts changing the half-mass radius by $\sim -0.5 \, \mathrm{dex}$, followed by a rapid recovery of its previous value. This implies a large mass of newborn stars is initially created at the centre, but also that a comparable mass is produced outside $r^{\star}_\mathrm{hm}$ afterwards. Therefore, quenching must proceed rapidly outwards and involve a large mass and a large spatial scale, both well above resolution. Moreover, their half-mass radii are $r^\star_\mathrm{hm} \gtrsim 1\,\mathrm{kpc}$, indicating a more extended star-formation region than in both our simulation and observed galaxies. The discrepancy may therefore probably be related to different physical modelling. Indeed, our resolution is comparable to the lowest one in the \textsc{Thesan} simulations. We note that extreme resolution values of a few tens of parsecs and hundreds of solar masses require a thorough evaluation of sub-grid processes, especially star formation, as some base assumptions like the Kennicutt--Schmidt law are not tested down to such scales. 

In principle, the gas flows that lead to compaction can be affected by stellar feedback. Stronger stellar feedback and larger velocity dispersion in the galaxies may disrupt the compaction described here until larger masses can be reached. However, different simulations with very different feedback models all describe a qualitatively similar central or bulge growth \citep{Ceverino2010,Genel2012,Mandelker2017,Oklopcic2017}. It seems that dense gas inflows at galaxy-scale are very resilient to the injection of energy and/or momentum. In particular, \citet{Ceverino2023} discuss the effect of two different feedback models on the compaction at relatively low redshifts ($z = 1-3$). The evolution of compaction and early quenching is similar in both cases, although stellar masses (and central densities) are a factor 2 lower in the strong feedback case. On the other hand, the addition of AGN feedback may affect the growth of massive galaxies, especially during their last re-expansion phase \citep{Lapiner2024}.

On the other hand, a careful comparison between observations and simulations involves several critical details that need to be addressed. In this context, it is essential to fully consider the observational procedures. We are currently addressing this issue in a forthcoming paper. It is not clear if an inverted size--mass relation should be directly observable due to the spread in the masses at which galaxies start and finish compaction. Drivers responsible for this spread should be explored in order to identify galaxy subsamples that may cleanly show the inversion, before a robust conclusion can be drawn from the observed size--mass relation. Indeed, although not a physical driver, our results show that redshift may indirectly play this role: the higher the redshift, the cleaner the inverted size--mass relation appears in our simulations. A large sample of galaxies in the aforementioned mass range and beyond $z \sim 8$ may provide clues on this issue. 

Another issue is the transformation of the physical size--mass relation into an observable one. Observers tackle the inverse problem through broadband SED fitting \citep[e.g.,][]{Roper2022,Morishita2024}, parameterising their models with the stellar mass and star formation history, and including the emission of ionized gas and dust absorption. The masses obtained are therefore subject to uncertainties and degeneration present in stellar evolutionary models, radiation transfer physics, etc. On the other hand, the sizes are measured from projected luminosity profiles, which are not easily deconvolved into 3D light distributions, and are subject to observational biases. A different (direct) approach is to construct galaxy images from the simulations, and compare directly the size--mass relations obtained from them. Biases can be included in the construction, and their effects on the relation explored. We will be presenting the results of such an analysis in a forthcoming paper.

Finally, an important characteristic of the proposed self-reinforced gas infall to the centre of galaxies is that it is self-triggered. We have shown that the accretion of this gas proceeds in the \citet{Bondi1944} regime, the flow being dependent of the square of the accreting mass. As far as the accreted mass overcomes that originally present in this region, the infall is dominated by the former and the process becomes self-reinforced. Indeed, we observe this change to occur because the inflowing matter is baryon-dominated, whereas the original mass in the central regions is dominated by dark matter. The switching between both regimes (from the prevalence of dark matter to that of baryons) is clearly seen in our simulations and supports our analysis of the physical mechanism behind compaction. It also explains why mass is its main driver. The \citet{Bondi1944} regime describes spherical accretion, in which angular momentum of the accreted material can be neglected. The ECE process may therefore constitute a continuation of the outer angular momentum loss process highlighted by \citet{Lapiner2024}. It may also be responsible for the formation of spheroidal structures with no or little net angular momentum, that would become the bulges of low-redshift galaxies, as proposed by \citet{Roper2022}. As the re-expansion phase begins and star formation moves to larger radii, angular momentum should play a more relevant role, thus changing or stopping the infall regime, and favouring the emergence of rotation-dominated structures such as discs. An investigation of the morphology and dynamics of the structures created during the ECE process is under way, and will be presented elsewhere.

In summary, our main conclusions are the following:

\begin{itemize}

\item
We have described the complete compaction process of galaxies since its onset to the late re-expansion, providing evidence for the wet compaction scenario put forth by previous works.

\item
Together with previous results, our findings suggest that this scenario is universal, operating at all redshifts.

\item 
The re-expansion ending compaction proceeds without central quenching at high redshift, contrary to its low-redshift analogue. This may be due to the higher gas density of galaxies at earlier epochs.

\item
We provided a physical explanation for the onset of wet compaction and its mass scale, based on Bondi--Hoyle accretion.
In addition, we introduced a new analytical expression, potentially valuable for use in semi-analytical models.

\end{itemize}

\begin{acknowledgements}

The authors acknowledge insightful comments by the anonymous referee, that helped to greatly improve the manuscript. PC, SP, LJP, and LAB acknowledge partial support from ANPCyT through grant PICT 2020-00582. This project has received funding from the European Union’s Horizon 2020 Research and Innovation Programme under the Marie Skłodowska-Curie grant agreement No 734374 (LACEGAL). DC is supported by research grant PID2021-122603NB-C21 funded by the Ministerio de Ciencia, Innovación y Universidades (MICIU/FEDER) and the research grant CNS2024-154550 funded by MICIU/AEI/10.13039/501100011033. The authors gratefully acknowledge the  Gauss Center for Supercomputing for funding this project by providing computing time on the GCS Supercomputer SuperMUC at Leibniz Supercomputing
Center (Project ID: pr92za). The authors thankfully acknowledge the computer resources at MareNostrum and the technical support provided by the Barcelona Supercomputing Center (RES-AECT-2020-3-0019). We acknowledge the publicly available programming language python (van Rossum 1995), including the numpy (Harris et al. 2020), astropy (Astropy Collaboration et al. 2013), matplotlib (Hunter 2007), scipy (Virtanen et al. 2020), and h5py (Collette 2013) packages.
\end{acknowledgements}

\section*{Data Availability}

The data underlying this article will be shared on reasonable request to the corresponding authors. 

\bibliographystyle{aa}

\end{document}